\documentstyle[12pt,aasms4]{article}

\received{October 4 1999}
\accepted{November 17 1999}

\begin{document}

\title{Correlation between the halo concentration $c$ and the virial mass 
$M_{vir}$ determined from X-ray clusters}

\author{Xiang-Ping Wu and Yan-Jie Xue}

\affil{Beijing Astronomical Observatory, Chinese Academy
                 of Sciences, Beijing 100012; 
       and National Astronomical Observatories, Chinese Academy
                 of Sciences, Beijing 100012; China}

\begin{abstract}
Numerical simulations of structure formation have suggested that there exists
a good correlation between the halo concentration $c$ (or the characteristic
density $\delta_c$) and the virial mass $M_{vir}$ for any virialized 
dark halo described by the Navarro, Frenk \& White (1995) density profile. 
In this {\sl Letter}, we present an observational determination of 
the $c$--$M_{vir}$ (or $\delta_c$--$M_{vir}$) relation in the mass range of 
$\sim10^{14}M_{\odot}< M_{vir} <\sim10^{16}M_{\odot}$
using a sample of 63 X-ray luminous clusters. The best-fit power law relation,
which is roughly independent of the values of $\Omega_M$ and 
$\Omega_{\Lambda}$, is $c\propto M_{vir}^{-0.5}$ or 
$\delta_c\propto M_{vir}^{-1.2}$, indicating 
$n\approx-0.7$ for a scale-free power spectrum of the primordial 
density fluctuations. We discuss the possible reasons for 
the conflict with the predictions by typical CDM models such as
SCDM, LCDM and OCDM.   
\end{abstract}

\keywords{cosmology: theory --- galaxies: clusters: general ---  
          X-rays: galaxies}

\section{Introduction}

High-resolution simulations of structure formation 
have suggested that the virialized dark matter halos 
with masses spanning several orders of magnitude should follow a universal
density profile (Navarro, Frenk \& White 1995; NFW)
\begin{equation}
\rho(r)=\frac{\rho_s}{(r/r_s)(1+r/r_s)^2}, 
\end{equation}
where $\rho_s$ and $r_s$ are the characteristic density and length, 
respectively.
The former is further related to the characteristic density $\delta_c$ 
and the critical density $\rho_c=(3H_0^2/8\pi G)Z(z)$ of the Universe 
at redshift $z$ through $\rho_s=\delta_c\rho_c$, in which
$Z(z)=(1+z)^3(\Omega_M/\Omega_M(z))$, and 
$\Omega_M$ is the cosmic density parameter.
Although the NFW profile was first obtained
in the SCDM model, subsequent numerical studies have shown that this
density profile is independent of mass, initial density fluctuation or
cosmology (e.g. Cole \& Lacey 1996; Navarro, Frenk \& White 1997; Eke, 
Navarro \& Frenk 1998; Jing 1999; etc.).  
The two free parameters in NFW profile can
be determined from the halo concentration $c=r_{vir}/r_s$ and the virial
mass $M_{vir}$ if the overdensity of the dark matter with respect to
the average value is $\Delta_c$,
\begin{eqnarray}
\delta_c=\frac{\Delta_c}{3}\frac{c^3}{\ln (1+c)-c/(1+c)};\\
r_s=\frac{1.626\times10^{-5}}{c}
    \left(\frac{M_{vir}}{M_{\odot}}\right)^{1/3}
    \left(\frac{\Delta_c}{200}\right)^{-1/3}	
    \left(\frac{1}{Z}\right)^{1/3} h^{-2/3}\; {\rm Mpc}.
\end{eqnarray}
It has been well established based on numerical simulations
that there exists a good correlation between the halo concentration $c$
(or the characteristic density $\delta_c$) 
and the virial mass $M_{vir}$ for any particular cosmology (e.g. Navarro,
Frenk \& White 1996, 1997; Salvador-Sol\'e, Solanes \& Manrique 1998; etc.). 
For example, given $\Delta_c=200$, in the mass range of 
$3\times10^{11}M_{\odot}\leq M_{vir}\leq 3\times10^{15}M_{\odot}$, 
the $c$-$M_{vir}$ relation can be well fitted by 
a power-law function: $c=a(M_{vir}/M_{\odot})^b$, with $(a,b)=(891,-0.14)$
and $(186,-0.10)$ for SCDM model 
($\Omega_M=1.0$, $\Omega_{\Lambda}=0$, $h=0.5$ and $\sigma_8=0.65$)
and $\Lambda$CDM model
($\Omega_M=0.25$, $\Omega_{\Lambda}=0.75$, $h=0.75$ and $\sigma_8=1.3$),
respectively. While the NFW profile and 
the $c$-$M_{vir}$ correlation
are only determined from simulations for halo masses as small as 
$M_{vir}=10^{11}M_{\odot}$ or as large as $M_{vir}=10^{15}M_{\odot}$,
there is no reason to believe that these results would not be valid
for halos which are one order of magnitude lower or higher in mass
(Burkert \& Silk 1999).

On the other hand, observationally we can only measure
the distribution of baryonic matter rather than that of dark halo. 
It is crucial to link numerical predictions with astronomical observations.  
In a fully virialized system, the distribution and motion
of baryonic matter can be determined
by the underlying gravitational potential of the dark halo,
if the velocity dispersion or temperature profile can be well measured. 
It is thus possible to work out the dynamical
properties of the dark matter halo such as $M_{vir}$, $r_s$, $r_{vir}$, $c$ or 
$\delta_c$ from the observed distribution of 
luminous matter in a gravitationally 
bound system. This will eventually allow us to examine observationally
whether there is a 
correlation between $c$ (or $\delta_c$) and $M_{vir}$. Finally,  
a comparison of the observationally determined and numerically simulated 
$c$--$M_{vir}$ (or $\delta_c$--$M_{vir}$) 
correlations may provide a useful way to distinguish
different cosmological models.
In this {\sl Letter}, we will make an attempt for the first time 
to derive the $c$--$M_{vir}$ and $\delta_c$--$M_{vir}$ 
correlations on cluster scales 
by fitting the observed X-ray surface brightness profiles of clusters 
to those predicted by the NFW profile as the cluster dark halos 
under the assumption of isothermality. Note that 
there is a striking similarity between the distribution of
intracluster gas tracing the dark halo of the NFW potential 
and the conventional $\beta$ model. Makino, Sasaki \& Suto (1998)
have explicitly shown that the NFW profile via isothermal hydrostatic 
equilibrium results in an analytic form of gas distribution
\begin{equation}
n_{gas}(r)=n_{gas}(0)e^{-\alpha}(1+r/r_s)^{\alpha/(r/r_s)},
\end{equation}
in which $\alpha=4\pi G\mu m_p\rho_s r_s^2/kT$, and $\mu=0.585$ denotes
the mean molecular weight. Briefly, our task is to determine  
the best-fit parameters of $\alpha$ and $r_s$ 
for an ensemble of X-ray clusters in terms of eq.(4),
and then derive the relevant parameters
$\rho_s$ and $\delta_c$  in conjunction with
the X-ray temperature $T$. Finally, we examine the $c$-$M_{vir}$ 
and $\delta_c$--$M_{vir}$ correlations 
by solving eqs.(2) and (3).

\section{Sample}

Essentially, we select our cluster sample from two ROSAT PSPC
cluster catalogs: the 36 high-luminosity clusters by Ettori \& Fabian 
(1999; EF) and the 45 nearby clusters by Mohr, Mathiesen \& Evrard 
(1999; MME). The EF clusters have high X-ray luminosity 
$L_x>10^{45}$ erg s$^{-1}$, and a great fraction of them also 
have relatively  high redshift $z\approx0.1$--$0.3$. 
The best-fit values of $\alpha$ and $r_s$ for all the 36 EF clusters
have been given by EF for a cosmological model of 
$H_0=50$ km s$^{-1}$ Mpc$^{-1}$ and $\Omega_M=1$. 
A conversion of $r_s$ into the values in different 
cosmological models should be properly made, if needed. 
The MME clusters are taken from an X-ray flux
limited sample, and thereby located at relatively small redshift.
We perform the $\chi^2$ fit of the observed X-ray surface 
brightness profiles of the MME clusters to the theoretical prediction
$S_x\propto \int n_{gas}^2 d\ell$ according to thermal 
bremsstrahlung, where the integral is made along the line of sight.
A cross-identification reveals that 16 clusters are listed in both
samples, for which our fitted values of $\alpha$ and $r_s$ are fairly
consistent with those obtained by EF.
We further require that the X-ray temperature should be 
observationally available, and we take the temperature data from
Wu, Xue \& Fang (1999; and references therein).  
The finally merged EF and MME sample
to be used in our following analysis contains 63 clusters. 


\section{The $c$-$M_{vir}$ and $\delta_c$-$M_{vir}$ correlations}

We confine ourselves to a flat cosmological model with 
$\Omega_M+\Omega_{\Lambda}=1$. In this case, the density contrast 
depends on the value of $\Omega_M$ and can be approximated by
(Eke et al. 1998) $\Delta_c=178\Omega_M(z)^{0.45}$, and
$Z(z)=(1+z)^2\{1+z\Omega_M+[(1+z)^{-2}-1]\Omega_{\Lambda}\}$ 
and $\Omega_M(z)=\Omega_M(1+z)^3/Z(z)$. For a given cosmological model 
$(\Omega_M,\Omega_{\Lambda})$, we convert our best-fit 
values of $\alpha$ and $r_s$ into $\delta_c$ and then obtain the concentration
parameter $c$ and the virial mass $M_{vir}$ 
by numerically solving eqs.(2) and (3).  We display in Fig.1 an example of
the resultant $c$ and $\delta_c$  versus $M_{vir}$ 
for $\Omega_M=0.3$ and $\Omega_{\Lambda}=0.7$.
It is apparent that
there exists a strong correlation between $c$ or $\delta_c$ and $M_{vir}$.
We employ the Monte-Carlo simulations and the $\chi^2$ fitting method
to obtain the best-fit $c$-$M_{vir}$ and $\delta_c$-$M_{vir}$ 
relations which are assumed to be a power-law. This enables us to
include the measurement uncertainties in both axes.
Note that for the EF clusters we have not accounted for the uncertainties 
arising from the fitted parameters of $\alpha$ and $r_s$
since EF provided no information about their error estimates.
The results for a set of cosmological models are listed  in Table 1.
Additionally, 
we have tried the orthogonal distance regression technique ODRPACK
(e.g. Feigelson \& Babu 1992) and reached a steeper power index. 
For example, in the case of $\Omega_M=0.3$ and $\Omega_{\Lambda}=0.7$ we find
$c=10^{14.47\pm1.03}(M_{vir}/M_{\odot})^{-0.92\pm0.07}$ and
$\delta_c=10^{33.15\pm2.28}(M_{vir}/M_{\odot})^{-1.97\pm0.15}$ (see Fig.1).
Although ODRPACK can simultaneously account for the scatters around 
$M_{vir}$ and $c$ (or $\delta_c$), the goodness of the fit has not been
improved for our problem in the sense that the reduced $\chi^2$ for 
both $c$-$M_{vir}$ and $\delta_c$-$M_{vir}$ relations are significantly 
larger than those obtained using the $\chi^2$ fitting method along
with Monte-Carlo simulations.
In Table 1 we have also given the power index for a scale-free power
spectrum of initial density fluctuations implied by our best-fit 
$\delta_c$-$M_{vir}$ relation:
$\delta_c\propto M_{vir}^{-(n+3)/2}$ (NFW), which yields $n\approx-0.7$.

\placefigure{fig1}

\begin{deluxetable}{lllll}
 \tablewidth{40pc}
 \scriptsize
\tablecaption{The best-fit $c$-$M_{vir}$ and $\delta_c$-$M_{vir}$ 
              relations}
\tablehead{
\colhead{$\Omega_M$} &  
\colhead{$\Omega_{\Lambda}$} & 
\colhead{$c$-$M_{vir}$ correlation} &
\colhead{$\delta_c$-$M_{vir}$ correlation} &
\colhead{$n$} }
\startdata
0.15 & 0.85 & $c=10^{8.17\pm0.58}(M_{vir}/M_{\odot})^{-0.50\pm0.04}$ &
     $\delta_c=10^{20.77\pm1.54}(M_{vir}/M_{\odot})^{-1.14\pm0.10}$ &
     $-0.72\pm0.20$ \nl
0.30 & 0.70 & $c=10^{8.23\pm0.63}(M_{vir}/M_{\odot})^{-0.51\pm0.04}$ &
     $\delta_c=10^{20.83\pm1.64}(M_{vir}/M_{\odot})^{-1.15\pm0.11}$ &
     $-0.70\pm0.22$ \nl
0.50 & 0.50 & $c=10^{8.29\pm0.69}(M_{vir}/M_{\odot})^{-0.51\pm0.05}$ &
     $\delta_c=10^{20.89\pm1.76}(M_{vir}/M_{\odot})^{-1.16\pm0.12}$ &
     $-0.68\pm0.24$ \nl
0.70 & 0.30 & $c=10^{8.32\pm0.74}(M_{vir}/M_{\odot})^{-0.52\pm0.05}$ &
     $\delta_c=10^{20.92\pm1.86}(M_{vir}/M_{\odot})^{-1.16\pm0.13}$ &
     $-0.68\pm0.26$ \nl
1.00 & 0.00 & $c=10^{8.35\pm0.81}(M_{vir}/M_{\odot})^{-0.53\pm0.06}$ &
     $\delta_c=10^{20.93\pm1.99}(M_{vir}/M_{\odot})^{-1.17\pm0.14}$ &
     $-0.66\pm0.28$ \nl
\enddata 
\end{deluxetable}

\section{Discussion and conclusions}

In a virialized system, the distribution and motion of 
galaxies and intracluster gas (if their self-gravity is negligible) 
are determined by the underlying gravitational potential
of the dark matter halo, provided that the velocity dispersion and
temperature profiles  are well measured.    
Therefore, one can probe the dynamical properties of
the dark matter halo, though it is invisible, 
by using optical/X-ray observations coupled with the 
hydrostatic equilibrium hypothesis. In this {\sl Letter}, we
have made an attempt to derive the halo concentration $c$,
the characteristic density $\delta_c$ and
the virial mass $M_{vir}$ for galaxy clusters characterized by the NFW profile
from the observed distribution and temperature of X-ray emitting gas.  
Although the correlation between $c$ (or $\delta_c$) 
and $M_{vir}$ has been well predicted from
numerical simulations, this is the first time to determine the 
$c$--$M_{vir}$ and $\delta_c$--$M_{vir}$
correlations making use of the real data from astronomical
observations.

The correlation between $c$ (or $\delta_c$) and $M_{vir}$ established 
in this {\it Letter} is applicable to massive halos in the mass range of 
$\sim10^{14}M_{\odot}<M_{vir}<\sim10^{16}M_{\odot}$. However, we notice
that the resultant slope ($\approx-0.5$) of the $\log c$--$\log M_{vir}$ 
relationship is significantly steeper 
while the spectrum ($n\approx-0.7\pm0.3$)
of the primordial density fluctuations is much flatter 
than the values predicted by typical CDM spectra for the mass halos with 
$3\times10^{11}M_{\odot}<M_{vir}<3\times10^{15}M_{\odot}$ 
(Burkert \& Silk 1999), especially on cluster scales where  
$n\approx-2$ (e.g. Henry \& Arnaud 1991; 
Mathiesen \& Evrard 1998; Donahue \& Voit 1999; Mahdavi 1999; etc.).
If our results are not a statistical fluke (Note the large
dispersion of the X-ray data points in Fig.1), 
the above conflict may imply that we should abandon 
at least one of our working hypotheses: 
(1)Intracluster gas is in hydrostatic equilibrium; 
(2)Intracluster gas has isothermal temperature profile; 
(3)The self-gravity of intracluster gas is considerably  small as compared
with the contribution of the dark matter halo;
(4)The NFW profile provides a precise description of the dark matter
distribution. Yet, further investigations should be made towards a 
robust constraint on the $c$--$M_{vir}$ and $\delta_c$--$M_{vir}$ relations
before we come to a detailed study of the possible reasons for
the reported discrepancy.

We note from Table 1 that our best-fit $c$--$M_{vir}$ and 
$\delta_c$--$M_{vir}$ relations and the constraints on $n$ are roughly 
independent of the cosmological models ($\Omega_m$ and $\Omega_{\Lambda}$).
This property will be significant for distinguishing different cosmological 
models when combined with high-resolution numerical simulations.

\acknowledgments
We gratefully acknowledge the valuable comments by an anonymous referee.
This work was supported by 
the National Science Foundation of China, under Grant No. 19725311.

\clearpage


 \begin{deluxetable}{llllll}
 \tablewidth{20pc}
 \scriptsize
\tablecaption{Cluster  sample$^*$}
 \tablehead{
\colhead{cluster} &  
\colhead{z} & 
\colhead{T (keV)} &
\colhead{$\alpha$} &
\colhead{$r{_s}$ (Mpc)}	}
\startdata
A85	& 0.0559 & 6.20$^{+0.40}_{-0.50}$ 
	& 8.465$^{+0.116}_{-0.116}$ & 0.360$^{+0.021}_{-0.021}$ \nl
A119  	& 0.0443 & 5.59$^{+0.27}_{-0.27}$ 
	& 12.30$^{+1.078}_{-1.078}$ & 2.585$^{+0.355}_{-0.355}$ \nl
A262	& 0.0169 & 2.41$^{+0.05}_{-0.05}$
	& 6.718$^{+0.113}_{-0.113}$ & 0.094$^{+0.008}_{-0.008}$ \nl 
A401	& 0.0737 & 8.00$^{+0.40}_{-0.40}$
	& 8.867$^{+0.092}_{-0.092}$ & 0.930$^{+0.027}_{-0.027}$ \nl
A426	& 0.0179 & 6.79$^{+0.12}_{-0.12}$
	& 8.084$^{+0.108}_{-0.108}$ & 0.181$^{+0.012}_{-0.012}$ \nl
A478	& 0.0881 & 6.90$^{+0.35}_{-0.35}$
	& 9.507$^{+0.107}_{-0.107}$ & 0.419$^{+0.018}_{-0.018}$ \nl 
A496	& 0.0325 & 4.13$^{+0.08}_{-0.08}$
	& 8.001$^{+0.118}_{-0.118}$ & 0.195$^{+0.014}_{-0.014}$ \nl
A520	& 0.2010 & 8.59$^{+0.93}_{-0.93}$
	& 11.35$^{+0.0}_{-0.0}$ & 1.70$^{+0.0}_{-0.0}$ \nl
A545	& 0.1530 & 5.50$^{+6.20}_{-1.10}$
	& 12.52$^{+0.0}_{-0.0}$ & 1.51$^{+0.0}_{-0.0}$ \nl
A586 	& 0.1710 & 6.61$^{+1.15}_{-0.96}$
	& 8.81$^{+0.0}_{-0.0}$ & 0.26$^{+0.0}_{-0.0}$ \nl
A644	& 0.0704 & 6.59$^{+0.17}_{-0.17}$
	& 8.771$^{+0.158}_{-0.158}$ & 1.175$^{+0.051}_{-0.051}$ \nl
A665	& 0.1816 & 8.26$^{+0.90}_{-0.90}$
	& 10.69$^{+0.0}_{-0.0}$ & 1.49$^{+0.0}_{-0.0}$ \nl
A754	& 0.0535 & 9.00$^{+0.50}_{-0.50}$
	& 10.33$^{+0.373}_{-0.373}$ & 0.774$^{+0.066}_{-0.066}$ \nl
A780	& 0.0552 & 3.57$^{+0.10}_{-0.10}$
	& 9.441$^{+0.372}_{-0.372}$ & 0.285$^{+0.046}_{-0.046}$ \nl
A1060	& 0.0126 & 3.24$^{+0.06}_{-0.06}$
	& 8.543$^{+0.199}_{-0.199}$ & 0.314$^{+0.018}_{-0.018}$ \nl
A1068 	& 0.1390 & 5.50$^{+0.90}_{-0.90}$
	& 9.72$^{+0.0}_{-0.0}$ & 0.42$^{+0.0}_{-0.0}$ \nl
A1367 	& 0.0214 & 3.50$^{+0.18}_{-0.18}$
	& 10.88$^{+0.696}_{-0.696}$ & 1.863$^{+0.179}_{-0.179}$ \nl
A1413	& 0.1427 & 8.85$^{+0.50}_{-0.50}$
	& 9.17$^{+0.0}_{0.0}$ & 0.57$^{+0.0}_{-0.0}$ \nl
A1651	& 0.0825 & 6.10$^{+0.20}_{-0.20}$
	& 9.082$^{+0.191}_{-0.563}$ & 0.563$^{+0.033}_{-0.033}$ \nl
A1656	& 0.0231 & 8.38$^{+0.34}_{-0.34}$
	& 22.36$^{+2.167}_{-2.167}$ & 3.471$^{+0.414}_{-0.414}$ \nl
A1689   & 0.1810 & 9.02$^{+0.40}_{-0.40}$
	& 10.93$^{+0.443}_{-0.443}$ & 0.715$^{+0.074}_{-0.074}$ \nl
A1763 	& 0.1870 & 8.98$^{+1.02}_{-0.84}$
	& 9.20$^{+0.0}_{-0.0}$ & 1.16$^{+0.0}_{-0.0}$ \nl
A1795 	& 0.0631 & 5.88$^{+0.14}_{-0.14}$
	& 10.04$^{+0.148}_{-0.148}$ & 0.462$^{+0.025}_{-0.025}$ \nl
A1835	& 0.2523 & 9.80$^{+1.40}_{-1.40}$
	& 10.22$^{+0.0}_{-0.0}$ & 0.32$^{+0.0}_{-0.0}$ \nl
A2029	& 0.0765 & 8.47$^{+0.41}_{-0.36}$
	& 9.198$^{+0.133}_{-0.133}$ & 0.390$^{+0.022}_{-0.022}$ \nl
A2052	& 0.0348 & 3.10$^{+0.20}_{-0.20}$
	& 8.358$^{+0.149}_{-0.149}$ & 0.204$^{+0.014}_{-0.014}$ \nl
A2063	& 0.0355 & 3.68$^{+0.11}_{-0.11}$
	& 8.194$^{+0.149}_{-0.149}$ & 0.362$^{+0.020}_{-0.020}$ \nl
A2142	& 0.0899 & 9.70$^{+1.30}_{-1.30}$
	& 9.446$^{+0.357}_{-0.357}$ & 0.636$^{+0.091}_{-0.091}$ \nl
A2163 	& 0.2030 & 14.69$^{+0.85}_{-0.85}$
	& 9.16$^{+0.0}_{-0.0}$ & 1.09$^{+0.0}_{-0.0}$ \nl
A2199 	& 0.0299 & 4.10$^{+0.08}_{-0.08}$
	& 9.131$^{+0.081}_{-0.081}$ & 0.315$^{+0.012}_{-0.012}$ \nl
A2204	& 0.1523 & 9.20$^{+1.50}_{-1.50}$
	& 8.633$^{+0.122}_{-0.122}$ & 0.236$^{+0.018}_{-0.018}$ \nl
A2218	& 0.1710 & 7.10$^{+0.20}_{-0.20}$
	& 10.32$^{+0.0}_{-0.0}$ & 0.99$^{+0.0}_{-0.0}$ \nl
A2219	& 0.2280 & 12.40$^{+0.50}_{-0.50}$
	& 11.51$^{+0.0}_{-0.0}$ & 1.59$^{+0.0}_{-0.0}$ \nl
A2244 	& 0.0970 & 8.47$^{+0.43}_{-0.42}$
	& 8.033$^{+0.223}_{-0.223}$ & 0.356$^{+0.037}_{-0.037}$ \nl
A2255	& 0.0808 & 7.30$^{+1.10}_{-1.70}$
	& 24.80$^{+6.635}_{-6.635}$ & 5.991$^{+1.982}_{-1.982}$ \nl
A2256	& 0.0581 & 7.51$^{+0.19}_{-0.19}$
	& 13.21$^{+0.655}_{-0.655}$ & 2.026$^{+0.180}_{-0.180}$ \nl
A2319   & 0.0559 & 9.12$^{+0.15}_{-0.15}$
	& 8.306$^{+0.143}_{-0.143}$ & 0.831$^{+0.041}_{-0.041}$ \nl
A2390	& 0.2279 & 11.10$^{+1.00}_{-1.00}$
	& 9.25$^{+0.0}_{-0.0}$ & 0.64$^{+0.0}_{-0.0}$ \nl
A2507	& 0.1960 & 9.40$^{+1.60}_{-1.20}$
	& 12.53$^{+0.0}_{-0.0}$ & 2.63$^{+0.0}_{-0.0}$ \nl
A2597	& 0.0852 & 4.40$^{+0.40}_{-0.70}$
	& 6.236$^{+0.783}_{-0.783}$ & 0.464$^{+0.129}_{-0.129}$ \nl
A2744 	& 0.3080 & 11.00$^{+0.50}_{0.50}$
	& 15.21$^{+0.0}_{-0.0}$ & 2.76$^{+0.0}_{-0.0}$ \nl
A3112 	& 0.0746 & 4.24$^{+0.24}_{-0.24}$
	& 8.412$^{+0.075}_{-0.075}$ & 0.209$^{+0.010}_{-0.010}$ \nl
A3158	& 0.0575 & 5.50$^{+0.30}_{-0.40}$
	& 10.28$^{+0.383}_{-0.383}$ & 1.093$^{+0.081}_{-0.081}$ \nl
A3266	& 0.0594 & 8.00$^{+0.30}_{-0.30}$
	& 14.11$^{+0.889}_{-0.889}$ & 2.438$^{+0.243}_{-0.243}$ \nl
A3391	& 0.0553 & 5.40$^{+0.60}_{-0.60}$
	& 7.203$^{+0.297}_{-0.297}$ & 0.579$^{+0.065}_{-0.065}$ \nl
A3526	& 0.0114 & 3.68$^{+0.06}_{-0.06}$
	& 6.985$^{+0.094}_{-0.094}$ & 0.057$^{+0.004}_{-0.004}$ \nl
A3532	& 0.0559 & 4.40$^{+4.70}_{-1.30}$
	& 9.146$^{+0.380}_{-0.380}$ & 0.934$^{+0.079}_{-0.079}$ \nl
A3558 	& 0.0475 & 5.12$^{+0.20}_{-0.20}$
	& 7.875$^{+0.205}_{-0.205}$ & 0.579$^{+0.040}_{-0.040}$ \nl
A3562 	& 0.0478 & 3.80$^{+0.50}_{-0.50}$
	& 6.938$^{+0.099}_{-0.099}$ & 0.378$^{+0.022}_{-0.022}$ \nl
A3571	& 0.0396 & 6.73$^{+0.17}_{-0.17}$
	& 9.253$^{+0.205}_{-0.205}$ & 0.659$^{+0.040}_{-0.040}$ \nl
A3667 	& 0.0566 & 7.0$^{+0.6}_{-0.6}$
	& 8.881$^{+0.158}_{-0.158}$ & 1.175$^{+0.051}_{-0.051}$ \nl
A4038   & 0.0302 & 3.30$^{+1.60}_{-0.80}$
	& 5.842$^{+0.082}_{-0.082}$ & 0.047$^{+0.004}_{-0.004}$ \nl
A4059	& 0.0478 & 3.97$^{+0.12}_{-0.12}$
	& 8.645$^{+0.198}_{-0.198}$ & 0.332$^{+0.023}_{-0.023}$ \nl
AWM7	& 0.0176 & 3.75$^{+0.09}_{-0.09}$
	& 8.756$^{+0.194}_{-0.194}$ & 0.357$^{+0.024}_{-0.024}$ \nl
Cygnus-A & 0.0570 & 6.50$^{+0.36}_{-0.36}$
	& 7.151$^{+0.071}_{-0.071}$ & 0.101$^{+0.008}_{-0.008}$ \nl
IRAS 09104 & 0.4420 & 8.50$^{+3.40}_{-3.40}$
	& 10.09$^{+0.0}_{-0.0}$ & 0.18$^{+0.0}_{-0.0}$ \nl
MKW3s 	& 0.0434 & 3.00$^{+0.30}_{-0.30}$
	& 8.422$^{+0.129}_{-0.129}$ & 0.219$^{+0.012}_{-0.012}$ \nl
MS1358 	& 0.3283 & 7.50$^{+4.30}_{-4.30}$
	& 14.29$^{+0.0}_{-0.0}$ & 1.48$^{+0.0}_{-0.0}$ \nl
MS2137  & 0.3130 & 4.37$^{+0.38}_{-0.72}$
	& 11.48$^{+0.0}_{-0.0}$ & 0.18$^{+0.0}_{-0.0}$ \nl
Ophia-A & 0.0280 & 9.10$^{+0.30}_{-0.30}$
	& 9.004$^{+0.558}_{-0.558}$ & 0.570$^{+0.093}_{-0.093}$ \nl
PKS 0745-19 & 0.1028 & 8.70$^{+1.60}_{-1.20}$
	& 8.930$^{+0.096}_{-0.096}$ & 0.270$^{+0.014}_{-0.014}$ \nl
Tria-A  & 0.0510 & 10.05$^{+0.69}_{-0.69}$
	& 9.277$^{+0.088}_{-0.088}$ & 1.042$^{+0.023}_{-0.023}$ \nl
ZW3146  & 0.2906 & 6.35$^{+0.37}_{-0.34}$
	& 10.26$^{+0.0}_{-0.0}$ & 0.19$^{+0.0}_{-0.0}$ \nl
 \enddata
 \tablenotetext{*}{This table is only available in electronic form.}
 \end{deluxetable}


\clearpage

\figcaption{The $c$--$M_{vir}$ and   $\delta_c$--$M_{vir}$ relations 
for 63 clusters in the case of 
$\Omega_M=0.3$ and $\Omega_{\Lambda}=0.7$.
The MME and EF clusters are represented
by the filled and open circles, respectively. Note that the error bars for
the EF clusters have only accounted for the temperature uncertainties.
The solid lines are the best $\chi^2$ fitted power law relations 
to the data sets,
while the dotted lines represent the results using the ODRPACK fitting method.
\label{fig1}}

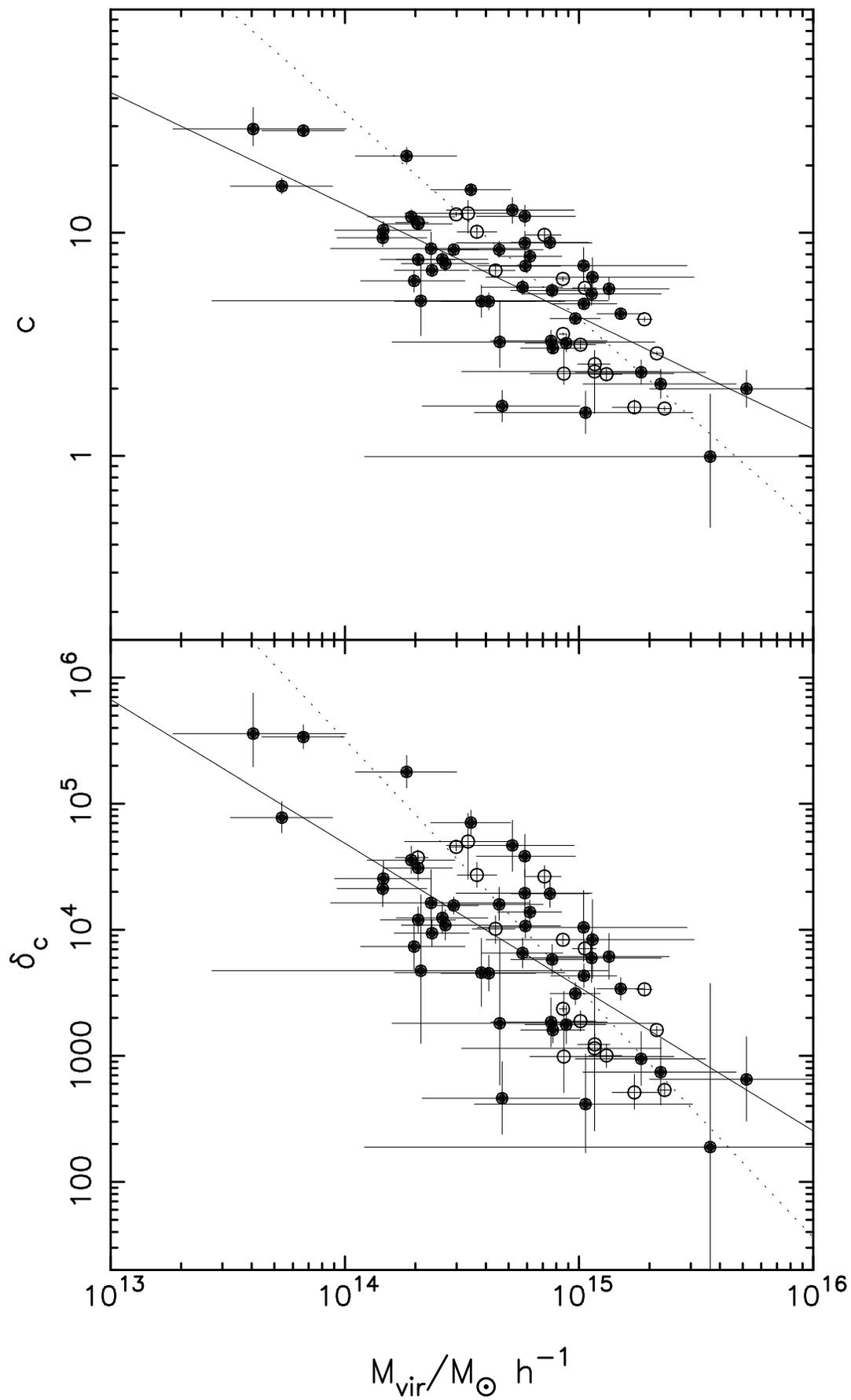
\end{document}